\documentclass[aps,prd,longbibliography ,twocolumn,nofootinbib,10pt]{revtex4-2}
\usepackage{graphicx, comment}
\usepackage{amsmath}
\usepackage{amsfonts}
\usepackage{graphicx}
\usepackage{dsfont}
\usepackage[german, english]{babel}
\usepackage{amssymb}
\usepackage{ifthen}
\usepackage{ulem}
\usepackage{color}
\usepackage{float}
\usepackage{physics }
\usepackage{hyperref}
\usepackage{bbold}
\usepackage{lipsum}
\usepackage{nccmath}
\usepackage{xcolor}

\hypersetup{colorlinks=true}

\newcommand{\br}{\boldsymbol{r}}

\newcommand{\bs}{\boldsymbol{s}}
\newcommand{\bt}{\boldsymbol{t}}

\newcommand{\bP}{\boldsymbol{P}}

\newcommand{\bzero}{\boldsymbol{0}}

\newcommand{\mycomment}[1]{}

\begin{document}

\title{Fractional Skyrme lines in ferroelectric barium titanate}
\author{Chris Halcrow}
\email{chalcrow@kth.se}
\author{Egor Babaev}
\affiliation{Department of Physics, KTH-Royal Institute of Technology, SE-10691, Stockholm, Sweden}
\date{\today}
 \begin{abstract}
     We predict a  topological defect in perfectly screened ferroelectric barium titanate which we call a skyrme line. These are line-like objects characterized by skyrmionic topological charge. As well as configurations with integer topological  charge, the charge density can split into well-localized  parts carrying a localized fraction of topological charge. We show that under certain conditions the fractional skyrme lines are stable. We discuss a mechanism to create fractional topological charge objects and investigate their stability.
 \end{abstract}

\maketitle


Skyrmions are topologically non-trivial defects originally proposed as a model of nuclei by Tony Skyrme \cite{skyrme1961}, though now various versions of skyrmions are more commonly studied in ferromagnets and other materials. They exist due to the topology of the system and their topological stability makes them promising candidates for various applications, with the prototypical application being high density memory storage \cite{fert2013skyrmions}.

Recently, ferroelectric materials were shown to be a host for non-trivial physics associated with topological defects. The most studied are lattices of polar vortices, predicted in \cite{naumov2004unusual, ponomareva2005low} and evidenced in \cite{yadav2016}. This motivated the theoretic \cite{nahas2015} and experimental \cite{das2019} searches for lattices with skyrmionic charge. The individual units of this lattice are sometimes called (polar) skyrmion ``bubbles". A bubble is a localised region of the material where the Polarization has an opposite direction to the background. The original bubbles \cite{kornev2004ultrathin, zhang2017nanoscale} were not topologically protected but skyrmion bubbles are \cite{hong2018blowing, yin2021nanoscale}. Evidence of localised skyrmion bubbles have been theoretically predicted in BaTiO$_3$ \cite{gonccalves2023} and strained PbTiO$_3$ \cite{pereira2019}.  Polar skyrmions and vortices have novel features such as chirality \cite{shao2023emergent} and local negative permittivity \cite{das2021local, yadav2019spatially}. Vortex-like Ising lines have also been simulated numerically \cite{stepkova2015}. 

In this paper, we show that a new type of defect in ferroelectrics is possible, which we coin a ferroelectric skyrme line. These share many properties with skyrmions, such as their topology and chirality, and can be thought of as skyrmions localized on ferroelectric domain walls. We report the existence of many non-trivial skyrme lines in barium titanate, despite not finding any stable skyrmions in our model. Configurations where skyrme topological charge is confined to a domain wall have attracted interest for a long time both in mathematical physics \cite{kudryavtsev1998skyrmions, kudryavtsev1999interactions, gudnason2014domain} and recently in other physical systems such as magnetic systems \cite{cheng2019magnetic, ross2023domain, nagase2021observation}
 and superconductors \cite{garaud2011topological,garaud2014domain}.
 
The especially interesting property of these defects is that they exhibit fractionalization of topological charge.
Traditionally topological charge is supposed to be an integer, represented as an integral over some topological charge density. However recently fractionalization, where the topological charge density is split into several stable configurations of localized fractions, has become of interest in a variety of models.
Fractional topological defects were searched for in various systems and a  fractional vortex was reported recently in superconductors \cite{iguchi2023superconducting}. Fractional skyrmions have been seen in condensed matter systems \cite{gao2020fractional, jena2022observation,nagase2021observation}
and mathematical physics
\cite{jaykka2012broken}.
The concept also applies to topological defects in higher dimensions
\cite{samoilenka2017fractional,samoilenka2020synthetic}.
In these examples, fractional defects exist as part of a larger object such as a lattice or integer-charged defect. Importantly,  we report that ferroelectrics allow topological line defects with unique fractional skyrme charge, which are themselves stable.

We study a Ginzburg-Landau-Devonshire model of barium titanate in the rhombohedral phase $(T<201$ K). The model can be written in terms of a Polarisation vector $\boldsymbol{P} = (P_1,P_2,P_3)$ and a symmetric strain tensor $u_{ij}$ which can be conveniently bundled into a 6-vector $e = (u_{11}, u_{22}, u_{33}, u_{23}, u_{13}, u_{12})$. The free energy density is given by
\begin{align}
\mathcal{F}  = \tfrac{1}{2}G_{abcd}\partial_aP_b \partial_cP_d + V(P) \nonumber \\ + \tfrac{1}{2}C_{\alpha\beta}e_\alpha e_\beta - q_{\alpha b c} e_\alpha P_bP_c \, .
\end{align}
The parameters are detailed in the supplementary material. The potential $V(P)$ in the rhombohedral phase has eight ground states which point in the directions of cube vertices: $\boldsymbol{P} \propto (1,1,1) $ etc. The strain tensor satisfies an addition compatibility constraint, which ensures there are no holes in the material. We have developed a new method to minimize the free energy while preserving the compatibility constraint. While minimizing for $P$, we project the energy-minimizing strain $e_\alpha = C^{-1}_{\alpha\beta}q_{\beta c d}P_cP_d$ onto a complete set of compatible functions. Since we can express the strain part of the free energy as an inner product, this is guaranteed to be the unique compatible energy-minimizing strain. More details can be found in the supplementary material. We ignore the electrostatic energy for simplicity and so our model is of perfectly screened barium titanate.

A skyrmion is a texture that exists and is stable due to the topology of the system. Usually, this topology is due to the structure of the field. In magnetic materials, the fundamental field is the magnetization: a field which takes values on the sphere $S^2$. Maps from $S^2$ to a plane with fixed boundary condition have non-trivial topology through the second homotopy group $\pi_2(S^2)$. It is this topology which makes the magnetic skyrmion stable. In ferroelectrics the order parameter field is the Polarisation $\boldsymbol{P} \in \mathbb{R}^3$, which has trivial topology. So a naive symmetry-based analysis would suggest that the field structure cannot support skyrmions. However, the point $\boldsymbol{P}=\boldsymbol{0}$ has very high energy and there is an energy cost for a configuration to contain this point.  This is why Bloch walls (which do not contain $\boldsymbol{P}=\boldsymbol{0}$) are often energetically favored over Ising walls (which contain $\boldsymbol{P} = \boldsymbol{0}$) in low temperature barium titanate \cite{Hlinka2006}. If $\boldsymbol{P}$ is never zero, the field can be thought of as $\boldsymbol{P} \in \mathbb{R}^3\setminus\{0\}$, which has non-trivial topology and can support skyrmions since $\pi_2( \mathbb{R}^3\setminus\{0\} ) = \mathbb{Z}$. 

\begin{figure}[h!]
	\begin{center}
            \includegraphics[width=\columnwidth]{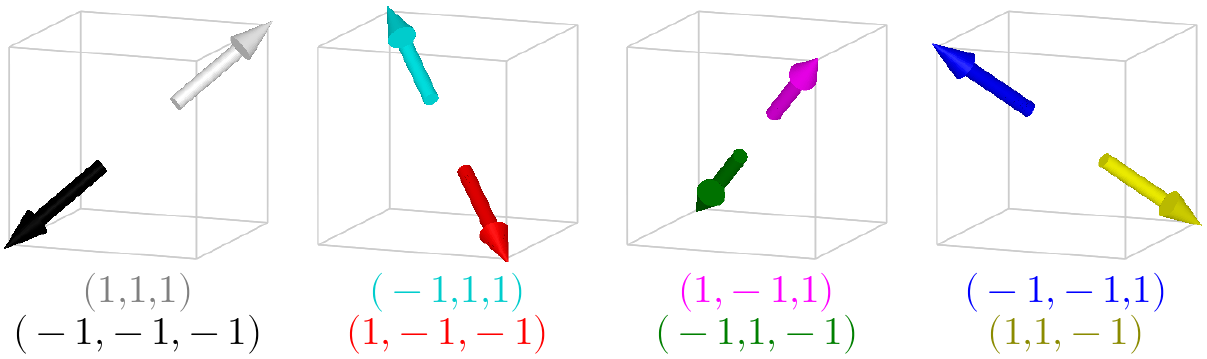}
		\includegraphics[width=0.9\columnwidth]{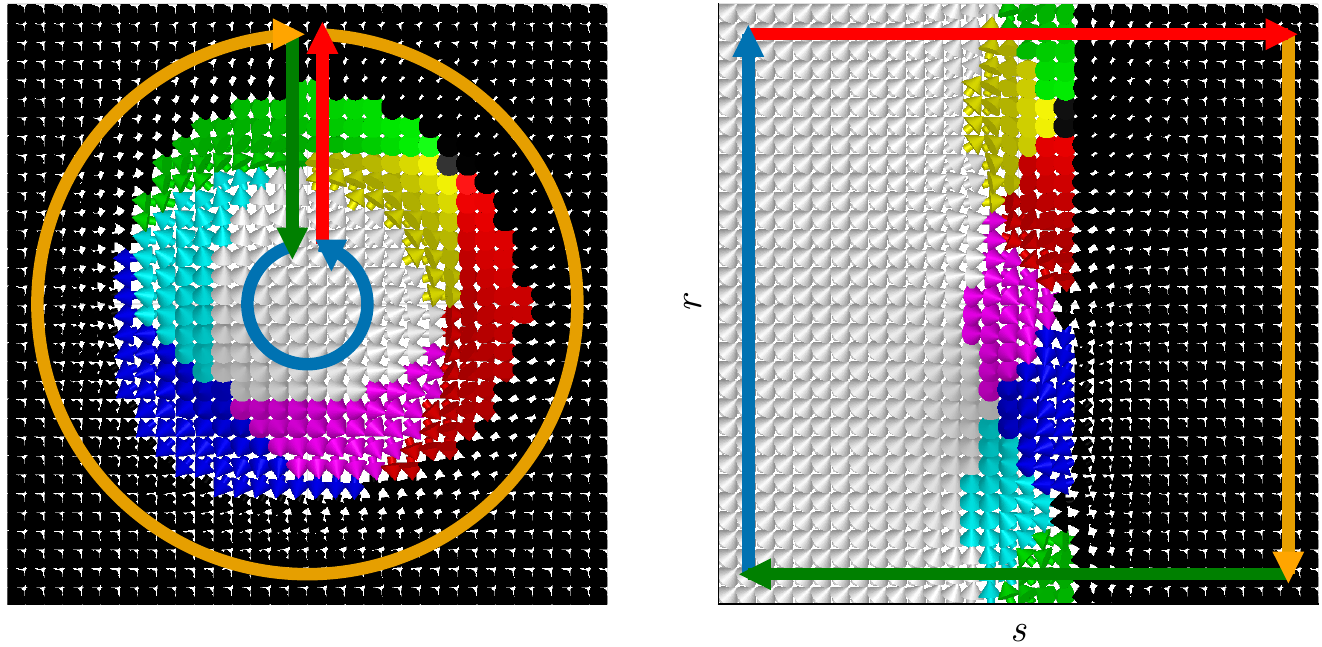}
		\caption{Plots of the Polarisation vector $\boldsymbol{P}$ for a skyrmion (left) and skyrme line (right). We find that ordinary skyrmions are unstable in barium titanate, but that skyrme lines can be stable. One can think of a skyrme line as an ``unwrapped" skyrmion. The long arrows demonstrate how a skyrmion is mapped onto a skyrme line. The short arrows, representing $\boldsymbol{P}$, are colored white, red, green, blue, teal, pink, yellow, and black when their nearest ground state is $P^V$ times $(1,1,1)$, $(-1,1,1)$, $(1,-1,1)$, $(1,1,-1)$, $(1,-1,-1)$, $(-1,1,-1)$, $(1,-1,1)$ and $(-1,-1,-1)$ respectively, visually represented in the top subfigure.}
		\label{fig:unwrap}
	\end{center}
\end{figure}

We find in barium titante a stable object which carries skyrmionic topological charge, which we call a skyrme line. These can be viewed as an``unwrapped skyrmion"   as shown in Fig. \ref{fig:unwrap}. The configurations are similar to domain walls but have a non-trivial structure along the wall. The skyrme lines lie on planes, embedded in $\mathbb{R}^3$. The free energy depends on the orientation of the 2D plane in the 3D material. We consider the plane spanned by two orthogonal vectors $\boldsymbol{s}$ and $\boldsymbol{r}$. First, consider a domain wall that connects two antipodal ground states $\pm\boldsymbol{P}^V$ along a direction $\bs$; the material's extent in the $\boldsymbol{s}$ direction, $L_s$, should be much larger than the width of the domain wall. We take periodic boundary conditions in the $r$-direction for simplicity and discuss more realistic boundary conditions in the next section. Overall, the boundary conditions are then
\begin{align}  \label{eq:boundary}
&\bP(\pm L_s,r) = \pm \bP^{V}\,, \quad \bP(s,L_r) = \bP(s,-L_r) \, .
\end{align}
The skyrme line seen in Fig. \ref{fig:unwrap} respects these boundary conditions. It is constructed using the initial configuration
\begin{equation} \label{eq:initial}
    P^\text{sk}_a(s,r;r_0) = \left|P^{V}\right|R_{ab}\begin{pmatrix} \cos(N_1r+r_0) \sin( f(s) )\\ \sin(N_1r+r_0) \sin( f(s) )\\ \cos( f(s)) \end{pmatrix}_b 
\end{equation}
where $R_{ab}$ is the rotation matrix taking $(0,0,1)$ to the boundary ground state, $(-1,-1,-1)$ in this case, and $f(-L_s) = 0, f(L_s) = N_2\pi$. The topological charge is equal to $N = N_1N_2$ and we have taken $N_1 = N_2 = 1$. Equation \eqref{eq:initial} is the unwrapped form of the standard `hedgehog' baby skyrmion \cite{piette1995multisolitons}. To have periodic boundary conditions $r$ must be a multiple of $\pi/L_r$. The parameter $r_0$ allows us to shift the skyrmion along the $r$-axis.

When the point $\bP = \bzero$ does not appear in a configuration we can construct the normalized Polarisation vector $\hat{\bP}$ and use it to calculate a topological charge, usually known as the skyrme charge:
\begin{equation} \label{eq:topcharge}
	N = \frac{1}{4\pi}\int \hat{\bP} \cdot  \partial_s \hat{\bP} \times \partial_r \hat{\bP}_c \, ds\,dr .
\end{equation}
The skyrme line in Fig. 1 has charge $N=1$. The charge is conserved provided that $\bP$ is never zero. If this does happen, the charge becomes undefined and the skyrme line collapses into a regular domain wall (as seen later in Fig. \ref{fig:linestring}).

\begin{figure}[h!]
	\begin{center}
		\includegraphics[width=\columnwidth]{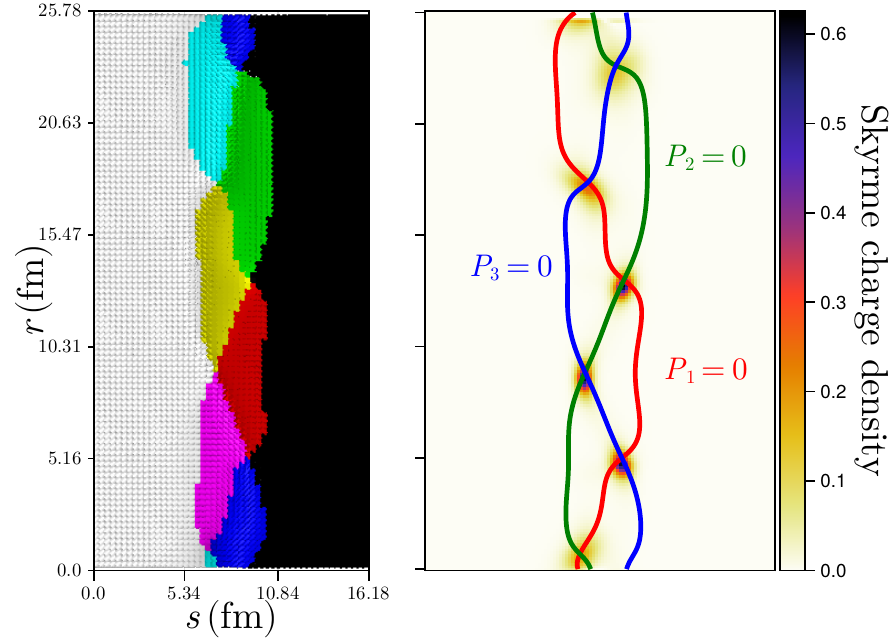}
		\caption{A numerically generated skyrme line plotted twice. We plot the Polarisation vector $\bP$ colored to reflect the closest vacuum (left) and the contours $P_i=0$ overlayed with the topological charge density (right). The charge is fractionalized: its density is mostly concentrated on the contour intersections.}
		\label{fig:ALine}
	\end{center}
\end{figure}

To show stability, we numerically relax the skyrme line \eqref{eq:initial} using a gradient flow \eqref{eq:gradflow}. We know that the allowed domain walls depend on the orientation of the wall in the material, and so we expect that the skyrme line stability depends on the plane orientation $(\bs, \br)$. We search over various plane orientations and find various stable configurations, including skyrme lines. One such configuration is plotted in Fig. \ref{fig:ALine}, with $\boldsymbol{s} = 1/\sqrt{3}(1,1,1)$ and $\boldsymbol{r} = 1/\sqrt{2}(0,1,-1)$. Note that the three contours $P_1=0, P_2=0$ and $P_3=0$ in Fig. \ref{fig:ALine} never touch. Their intersection would correspond to the point $\boldsymbol{P}=\boldsymbol{0}$, which has very high energy. One can only ``unknot" the contours by passing through the point. This energy barrier generates an outward pressue on the skyrme line. Conversely, the gradient energy $G_{abcd}\partial_aP_b\partial_cP_d$ is minimised when the line collapses into a simple domain wall and so encourages the line to shrink. The balance between these two forces stabilizes the Skyrme line.

The topological charge deserves careful study. We observe what can be interpreted as topological charge fractionalization in this system. Namely the topological charge density  is equally concentrated at the six points, where two of the three contours $P_i=0$ intersect. At these intersections, two of the $\bP$ components are zero and so the Polarisation points along a Cartesian axis. Now consider a loop around an intersection, which is in target- (or $\bP$-) space and encircles an axis. We call the orientation of this loop the chirality of the intersection. If the direction is anti-clockwise, the chirality is positive and vice-versa. Each positive chirality intersection contributes $+1/6$ to the charge and each negative chirality loop $-1/6$. In Fig. \ref{fig:ALine} the skyrmion contains six positive chirality intersections, and so has charge $N=1$. 
More formal arguments, using different language, were recently made for a magnetic system in \cite{Filipp}. The intersections are extrema of the potential and so each charge $1/6$ skyrmion has the same mathematical structure as their ``non-abelian vortices", which are topologically stable due to an energetics-motivated puncturing of the target manifold.


We have seen that the topological charge fractionalizes into sixths and will now show that we can construct fractional skyrme lines by adjusting the boundary conditions of the system. We again construct a domain wall connecting two ground states $\pm \boldsymbol{P}^V$. In barium titante, there are several energy-degenerate domain walls that connect the ground states. We suppose that one type of domain wall $P_{W+}(s)$ is present at one side of the material, $L_R$, and another $P_{W-}(s)$ at the other side $-L_R$. We expect this situation to occur when the system is annealing. Since there is no energetic reason for one wall to be preferred, both will form in different regions and our system describes what will happen between these regions. The boundary conditions are now
\begin{align}  \label{eq:boundary2}
&P(\pm L_s,r) = \pm \bP^{V}\,\\
&P(s,-L_r) = P_{W-}(s) \, , \quad P(s,L_r) = P_{W+}(s) \, ,
\end{align}
where $P_{W+}$ and $P_{W-}$ are genuine 1D domain wall solutions. These can be found in arbitrary orientations following the methods developed in \cite{halcrow2023ferroelectric}. We then generate an initial configuration that satisfies the boundary conditions \eqref{eq:boundary2}, of the form
\begin{equation} \label{eq:interp}
    \bP(s,r) = \bP_{W-}(s)g(r) + \bP_{W+}(s)(1-g(r)) \, ,
\end{equation}
with $g(L_r) = 0$ and $g(-L_r) = 1$.

\begin{figure}[h!]
	\begin{center}
		\includegraphics[width=\columnwidth]{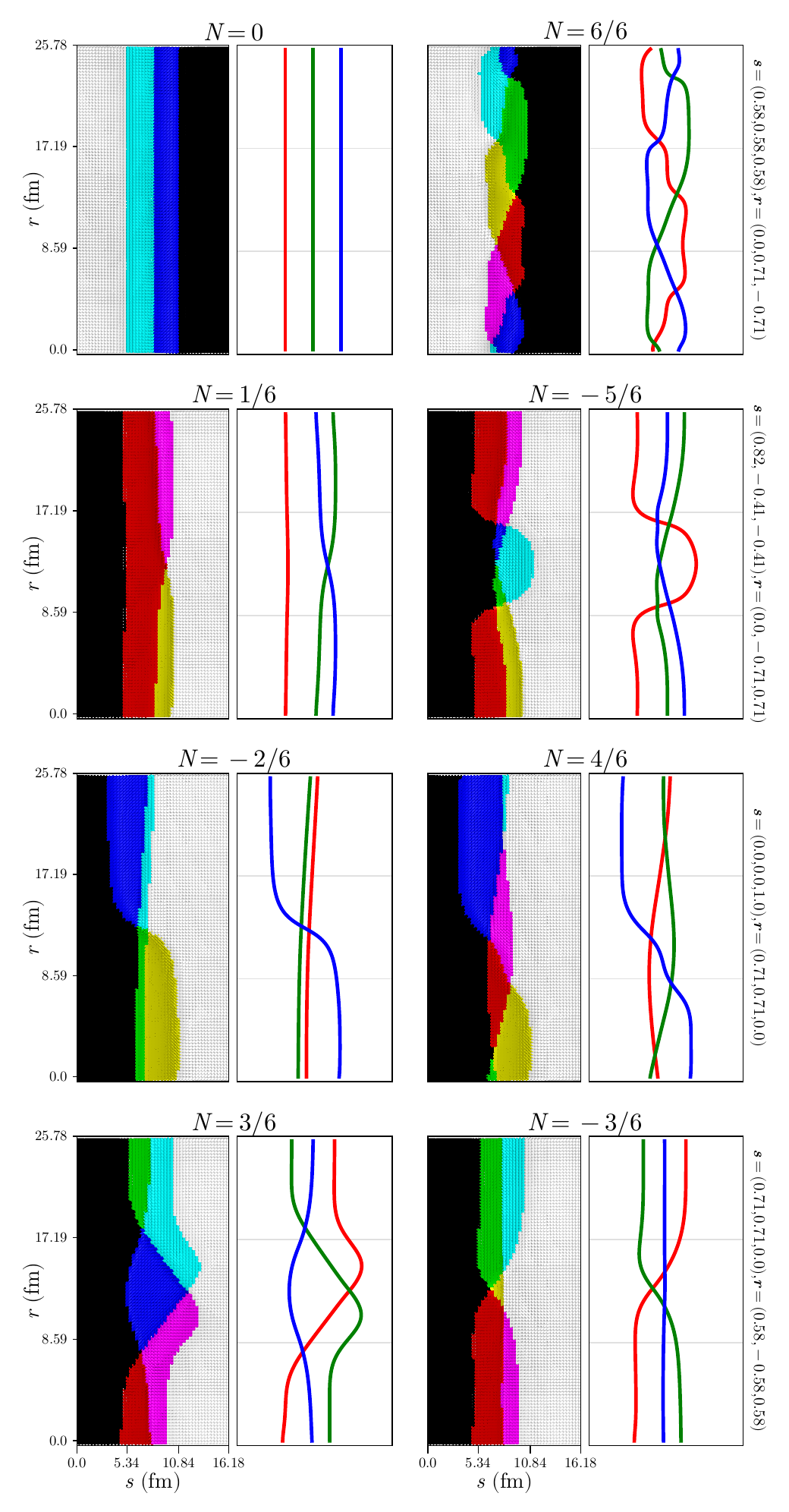}
		\caption{Fractional skyrme lines with topological charge $n/6$, which are stable in a variety of planes spanned by $\boldsymbol{s}$ and $\boldsymbol{r}$. Skyrme lines with the same boundary data have relative topological charge $1$.}
		\label{fig:lotsawalls}
	\end{center}
\end{figure}

We now apply gradient flow \eqref{eq:gradflow} to the initial data \eqref{eq:interp}. By using different $P_{W\pm}$ and different $g(r)$s, we can generate a zoo of skyrme lines.  Solutions with absolute topological charge $n/6, n \in [1,6]$ are plotted in Fig. \ref{fig:lotsawalls}. All configurations here have fixed boundary conditions. There are four different sets of boundary conditions, and for each we plot two configurations separated by one unit of charge. The fact that both are stable suggests that there is an energy barrier due to their different topological charges. We calculate the topological charge by simply counting the contour intersections, with chirality. Given the order of contours at the top and bottom of the box, the charge is fixed to be some fraction, up to an integer. Hence the fractional charge is due to the structure of the domain wall solutions at each side of the system. We verify this simple counting by calculating the topological charge numerically using \eqref{eq:topcharge}. For each skyrme line with charge $N$, there is an energy-degenerate partner with negative skyrmionic charge $-N$. This can be generated by applying a reflection to $\boldsymbol{P}$, across the plane with normal $\boldsymbol{P}^V \times \boldsymbol{s}$: a symmetry of the system.

We can better probe the wall stability by then relaxing the boundary condition to be open. That is, take $\partial_r P(s, \pm L_r) = 0$. When this is done, all the walls in Fig. \ref{fig:lotsawalls} are stable except the 4/6-charge wall, which ejects two intersections to become a 2/6-charge wall. Overall, we have found stable, localized skyrme lines with various fractional topological charges.

The skyrme lines discussed here have some similarities and differences with other objects in the literature. In \cite{nahas2015}, the authors found a lattice with skyrmionic charge $1$ per unit cell. Like ours, the charge fractionalizes and is concentrated at the intersection of contours $P_i=0$. Unlike ours, the lattice is only stable due to the presence of a nanowire. Recently, the existence of individual antiskyrmions (similar to the configuration shown in Fig. 1 left) in Barium-titanate at $T=0$K has been reported \cite{gonccalves2023}. Despite searching, we do not find such stable configurations in our simulations. Finally, note that the charge $N=\pm1/2$ skyrme lines have the same topological structure as merons.


To test the stability of the skyrme lines we construct a string of configurations joining the skyrme line to the ground state domain wall. We then flow the entire string of configurations while keeping the distance between the configurations fixed. This is the simplified string method, which has been applied to chemical reactions \cite{weinan2007}, superconductors \cite{benfenati2020} and ferroelectrics \cite{halcrow2023ferroelectric}. The process generates the minimal energy path in configuration space joining the skyrme line to the ground state.

\begin{figure}[h!]
		\begin{center}
		\includegraphics[width=\columnwidth]{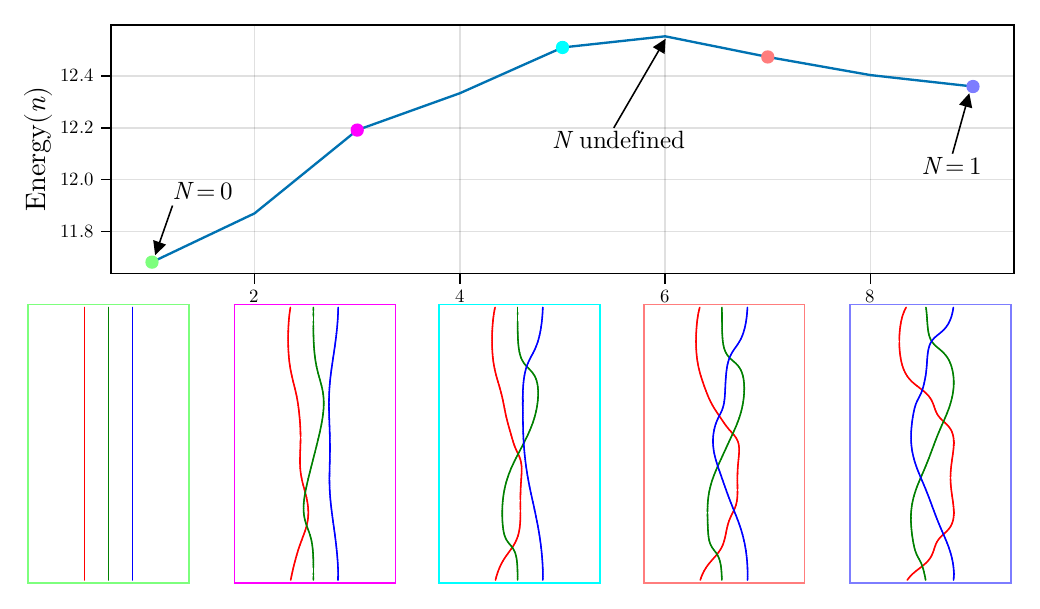}
        \includegraphics[width=\columnwidth]{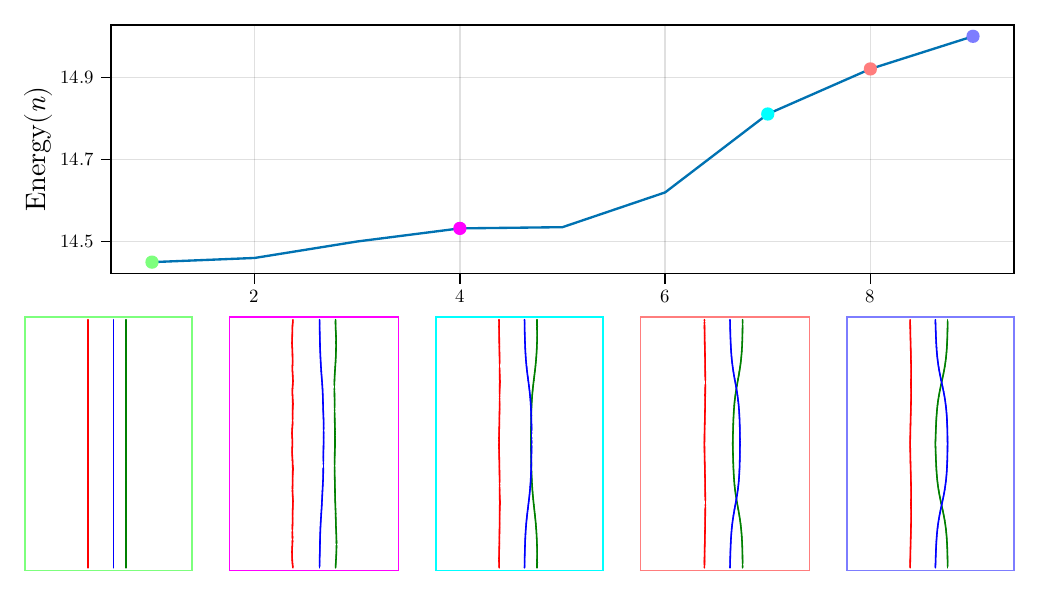}
		\caption{Strings of configurations. Top: A simple domain wall (left) joined to a charge 1 skyrme line (right) with $\boldsymbol{s} = (1,1,1)/\sqrt{3}$ and $\boldsymbol{r} = (0,1,-1)/\sqrt{2}$. The string passes through a high-energy saddle point configuration where the charge becomes ill-defined. Bottom: A simple domain wall (left) joined to a charge 0 skyrme line made from $+1/6$ and $-1/6$ skyrme lines (right), with $\boldsymbol{s} = (1,1,1)/\sqrt{3}$ and $\boldsymbol{r} = (0,1,-1)/\sqrt{2}$. There is no energy barrier between the configurations since they have the same topological charge.  We plot the energy for each configuration on the string, and the contour lines for several of the configurations. The energy is in dimensionless units. These can be converted to Joules through the factor $\sqrt{G_{11}^3(P^V)^4/A_{11}}$.} 
  		\label{fig:linestring}
    \end{center}
\end{figure}

We show the results for two simulations in Fig. \ref{fig:linestring}.
The top plot displays a string of configurations joining a skyrme line, with topological charge one, to a simple domain wall with topological charge zero. 
Each end of the string is a local minimum and there is a peak between the solutions which represents the barrier separating the two.
Near the peak of the barrier the three contours meet, representing the point $\boldsymbol{P} = \boldsymbol{0}$, where the topological charge becomes ill-defined.
Hence the energy barrier seen in Fig. \ref{fig:linestring} represents the energy cost of passing through this point and breaking the topology.
The bottom plot displays a second string that joins two configurations with the same topological charge. The right-most line is constructed from a $1/6$ and a $-1/6$ line.
Here, there is no topological barrier and so there is no energy barrier to the ground state domain wall.
This example demonstrates that topological features are a key input for the stability of ferroelectric defects.

\begin{figure}[h!]
	\begin{center}
		\includegraphics[width=\columnwidth]{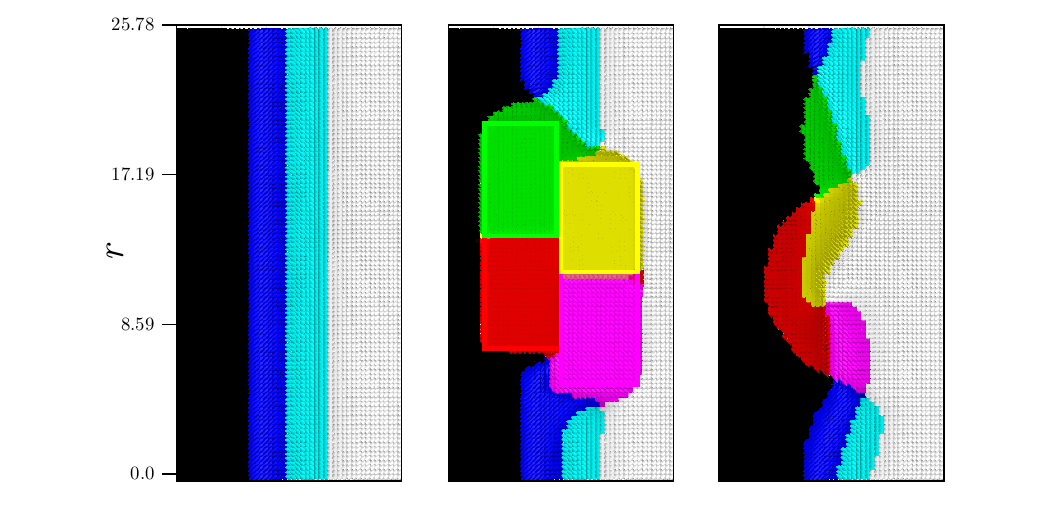}
		\includegraphics[width=\columnwidth]{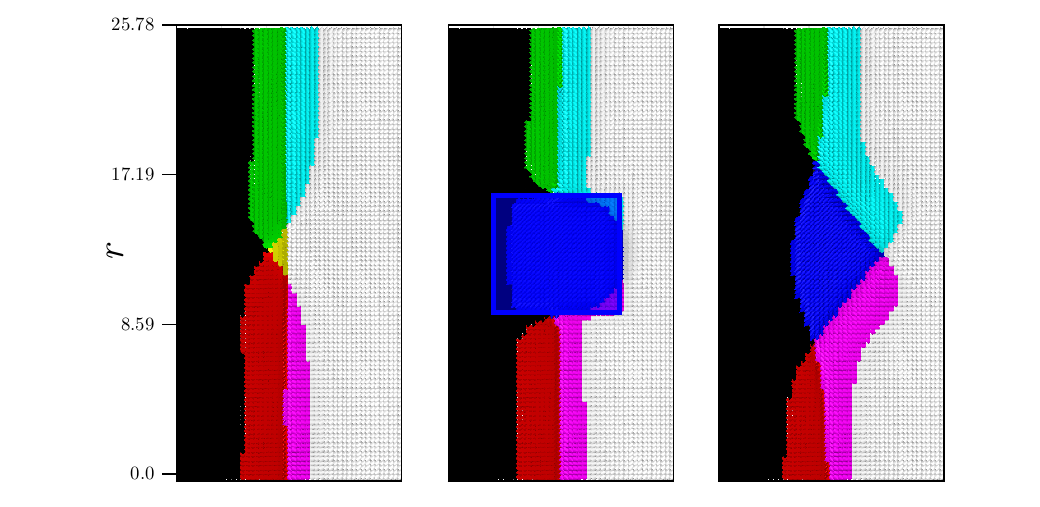}
		\caption{Making a charge $1$ skyrme line (top) and switching a $N=-1/2$ line to a $N=+1/2$ line (right) using external electric fields. The color of each stencil represents the direction of the applied external field.}
		\label{fig:Eext}
	\end{center}
\end{figure}

The skyrme lines that we found can be created and manipulated using external electric fields. 
First, one can create a whole skyrme line from a configuration with zero topological charge. 
The process requires a complicated stencil; we use a four-part stencil. Each part of the electric field stencil had a strength $|E| = $0.3 MeV/m$^2$. Each square was $3\times 6$nm large and point in the direction indicated by the colour in Fig. \ref{fig:Eext}. The colour scheme is visually represented in Fig. \ref{fig:unwrap}. The electric field was applied until the system reached equilibrium (less than $1 s$, in units of $\tau$. See eq \eqref{eq:gradflow}), then turned off. This creates enough energy to overcome the $\bP = \bzero$ energy barrier and a skyrme line with topological charge $N=1$ is created. 

Since each part of the previous stencil points in a different direction, engineering such an external field is a challenge. A simpler process, with a simpler stencil, can be seen at the bottom of Fig. \ref{fig:Eext}. 
Here, we switch from a $N=+1/2$ line to a $N=-1/2$ line using a one-piece stencil. 
A $6 $nm$\times 6$nm electric field of strength 0.3 MeV/m$^2$ was applied until equilibrium was achieved (less than $1 s$, in units of $\tau$), then turned off.
In the process, three positive-chirality contour intersections are turned into three negative-chirality ones.


We have considered a new type of topological defect in ferroelectric barium titanate, which we call a skyrme line. Their stability depends on a topological charge which is protected by the high energy cost of the point $\boldsymbol{P} = \boldsymbol{0}$. We found stable skyrme lines and studied their stability, creation, and switching. A unique feature of ferroelectric skyrme lines is that they can exist with fractional topological charge and these structures should appear naturally in an annealed sample. We found all examples of possible fractional skyrme lines $N=n/6$ in the rhombohedral phase.

In this work, we have ignored the electrostatic energy contribution. Physically, we have modelled perfectly screened barium titante. This approximation is most reasonable for neutral domain walls, when $\boldsymbol{s}\cdot \bP = 0$, such as the charge $N = \pm 1/2$ walls studied here. These are the skyrme line most likely to exist in real barium titante. To include the electrostatic energy, one can add a non-local term which is numerically expensive. Including also raises difficult theoretical questions of regularization, especially for non-periodic boundary conditions \cite{hu1997computer}.

We discussed how to switch domain walls using a very simple mechanism. The robustness and manipulability of these objects suggest that fractional  skyrme lines might be useful objects for data storage devices. The $+1/2$ and $-1/2$ skyrmions have different chirality, and this property might be determinable using 4D-STEM experiments \cite{shao2023emergent}.

\begin{acknowledgments}
We thank Katia Gallo, Anton Talkachov, Mats Barkman and Albert Samoilenka for useful discussions. CH is supported by the Carl Trygger Foundation through the grant CTS 20:25. This work is supported by the Swedish Research Council Grants 2016-06122 and 2022-04763
and by the Knut and Alice Wallenberg Foundation through the Wallenberg Center for Quantum Technology (WACQT).
\end{acknowledgments}


%

\setcounter{equation}{0}
\setcounter{figure}{0}
\setcounter{table}{0}
\setcounter{section}{0}
\makeatletter
\renewcommand{\theequation}{S\arabic{equation}}
\renewcommand{\thefigure}{S\arabic{figure}}
\pagebreak

\onecolumngrid
\vskip 0.4cm
\begin{center}
\textbf{\large Supplemental Material}
\end{center}
\vskip 0.4cm
\twocolumngrid
\section{Ginzburg-Landau Devonshire model of barium titante}

The Ginzburg-Landau-Devonshire model of barium titanate can be written in terms of a Polarisation vector $\bP = (P_1, P_2, P_3)$ and a 6-vector containing the strains $e = (u_{11},u_{22},u_{33},u_{23},u_{13},u_{12})$. The free energy is given by
\begin{align} \label{eq:freeenergy}
&F = \int \tfrac{1}{2}G_{abcd}\partial_aP_b \partial_cP_d + V(P)\, d^3x + \tfrac{1}{2}|| e - QPP ||_C \\
&Q_{\alpha b c} = C^{-1}_{\alpha \beta}q_{\beta b c} \, .
\end{align}
where  $(QPP)_\alpha = Q_{\alpha b c}P_b P_c$, the potential energy is given by 
\begin{align*}
V = &A_{ab}P_aP_b + A_{abcd}P_aP_bP_cP_d + A_{abcdef}P_aP_bP_cP_dP_eP_f
\end{align*}
and the distance $|| \cdot ||_C$ is measured with respect to the inner product
\begin{equation} \label{eq:innerC}
	\langle e^{(1)}, e^{(2)} \rangle_C = \int C_{\alpha \beta}e^{(1)}_\alpha e^{(2)}_\beta d^3x \, .
\end{equation}
We bundle the parameters into tensors as follows
\begin{align*}
    &A_{ij} = \delta_{ij}\alpha_1 \\
    &A_{ijkl} = \left(\alpha_{11} - \tfrac{1}{2}\alpha_{12} \right) \delta_{ijkl} + \tfrac{\alpha_{12}}{6} \left(\delta_{ij}\delta_{kl} + \delta_{ik}\delta_{jl} + \delta_{il}\delta_{jk} \right) \\
    &A_{ijklmn} = \left( \alpha_{111}-\alpha_{112} + \tfrac{1}{3}\alpha_{123}\right) \delta_{ijklmn} \\
     &\qquad + \tfrac{1}{6}a_{123}\delta_{ij}\delta_{kl}\delta_{mn} \\
     &\qquad + \left(\tfrac{1}{15}a_{112} - \tfrac{1}{30}a_{123} \right)\left( \delta_{ij}\delta_{klmn} + \text{14 perms} \right) \, .
\end{align*}
The tensors $C$ and $G$ have the same decomposition. We give it for $C$,
\begin{align*}
    C_{ijkl} &= \left(C_{11} - C_{12} - 2C_{44}\right)\delta_{ijkl} + C_{12}\delta_{ij}\delta_{kl} \\
    &+ C_{44}\left(\delta_{ik}\delta_{jl} + \delta_{il}\delta_{jk} \right) \, ,
\end{align*}
while $q$ and $Q$ are similar but slightly modified,
\begin{align*}
    q_{ijkl} &= \left(q_{11} - q_{12} - q_{44}\right)\delta_{ijkl} + q_{12}\delta_{ij}\delta_{kl} \\
    &+ \tfrac{1}{2}q_{44}\left(\delta_{ik}\delta_{jl} + \delta_{il}\delta_{jk} \right) \, .
\end{align*}
The coefficients can be found in Table \ref{tab:constants}, and are taken from using parameters from \cite{buessem1966, bell1984, Hlinka2006}.

\begin{table}[b]
	\begin{tabular}{l | c | l }
 Const. & Value & Units \\ \hline
		$\alpha_1$ & $3.34 (T-381)$ & $10^5$ JmC$^{-2}$\\
  $\alpha_{11}$ & $4.69(T-393) - 202$ & $10^6$ Jm$^5$C$^{-4}$ \\
  $\alpha_{12}$ & 3.23 & $10^8$ Jm$^5$C$^{-4}$ \\
  $\alpha_{111}$ & $-55.2(T-393) + 2760$ & $10^6$ Jm$^9$C$^{-6}$ \\
  $\alpha_{123}$ & 4.91 & $10^9$ Jm$^9$C$^{-6}$  \\
  $\alpha_{112}$ & 4.47 & $10^9$ Jm$^9$C$^{-6}$  \\ \hline
  $C_{11}$ & 2.75 & $10^{11}$ Jm$^{-3}$ \\
  $C_{12}$ & 1.79 & $10^{11}$ Jm$^{-3}$ \\
  $C_{44}$ & 5.43 & $10^{10}$ Jm$^{-3}$ \\ \hline
  $q_{11}$ & 1.42 & $10^{10}$ JmC$^{-2}$ \\
  $q_{12}$ & -7.4 & $10^{8}$ JmC$^{-2}$ \\
  $q_{44}$ & 1.57 & $10^{9}$ JmC$^{-2}$ \\ \hline
  $G_{11}$ & 51 & $10^{-11}$ Jm$^3$ C$^{-2}$ \\
  $G_{12}$ & -2 & $10^{-11}$ Jm$^3$ C$^{-2}$ \\
  $G_{44}$ & 2 & $10^{-11}$ Jm$^3$ C$^{-2}$ \\ 
	\end{tabular} 
	\caption{The material constants used in this paper.} \label{tab:constants}
\end{table}

The strain tensor must satisfy the compatibility condition 
\begin{equation}
	\epsilon_{abc}\epsilon_{def}\partial_b\partial_e u_{cf}=0\, ,
\end{equation}
which ensures that the strain is physically realizable and the material contains no holes. We'll focus on a two-dimensional plane, embedded in the three-dimensional material. Denote the spanning vectors of the plane as $\bs$ and $\br$, and the orthogonal vector $\bt = \bs \times \br$. Our new coordinates are then $(s,r,t)$. In Voigt-notation, the compatibility condition becomes
\begin{equation} \label{eq:econstraint}
	\frac{\partial^2 e_{1}}{\partial r^2} - 2  \frac{\partial^2 e_{6}}{\partial s\partial r} +  \frac{\partial^2 e_{2}}{\partial s^2} = 0 \, .
\end{equation}
The other strains $e_3,e_4,e_5$ are constant, respecting the trivial behavior in the $t$-direction. We'll use a rectangular lattice so we can write the solution of \eqref{eq:econstraint} as a Fourier Series, with constraints on the Fourier coefficients. There are eight solutions for each $(p,q)$-frequency pair:
\begin{align}
	e^{pq}_{1} =& a_{1} c_{ps}c_{qr} + b_{1} c_{ps}s_{qr} + c_{1}s_{ps}c_{qr} + d_{1}s_{ps}s_{qr} \\
	e^{pq}_{2} =& a_{2} c_{ps}c_{qr} + b_{2} c_{ps}s_{qr} + c_{2}s_{ps}c_{qr} + d_{2}s_{ps}s_{qr} \\
	e^{pq}_{6} =& a_{6} c_{ps}c_{qr} + b_{6} c_{ps}s_{qr} + c_{6}s_{ps}c_{qr} + d_{6}s_{ps}s_{qr}
\end{align}
where $s$ and $c$ represent sine and cosine, the arguments $p$ and $q$ depend on the boundary conditions and the coefficients of $e_6$ are fixed by \eqref{eq:econstraint} to be
\begin{align}
	a_6 &= -\frac{1}{2}\left( \tfrac{q}{p} d_1 + \tfrac{p}{q} d_2 \right),\quad b_6 = \frac{1}{2}\left( \tfrac{q}{p} c_1 + \tfrac{p}{q} c_2 \right) \\
	c_6 &= \frac{1}{2}\left( \tfrac{q}{p} b_1 + \tfrac{p}{q} b_2 \right), \quad d_6 = -\frac{1}{2}\left( \tfrac{q}{p} a_1 + \tfrac{p}{q} a_2 \right) \, .
\end{align}
Hence, we can write any compatible strain tensor as
\begin{align} \label{eq:easA}
	e_{\alpha} &= \sum_{p,q,a} A^{pqa}_\alpha e^{pqa}_\alpha \quad \text{for }\alpha = 1,2,6 \\
	e_{\alpha} &= A_\alpha \quad\quad\quad\quad \text{for }\alpha = 3,4,5 \, ,
\end{align}
where $a = \{1,\ldots,8\}$ enumerates the eight solutions.

The strain tensor which minimizes the free energy \eqref{eq:freeenergy} is the one that minimizes  $||e - QPP||_C$. Since we have a basis for all compatible strains, we can write this tensor explicitly using the Best Approximation theorem. To do so, we need a basis that is orthogonal with respect to \eqref{eq:innerC}. The choice of trigonometric functions ensures that terms with different frequencies $p$ and $q$ are automatically orthogonal. We are left to orthogonalize the eight solutions at each frequency, which we can do analytically using the Gram-Schmidt process. Denote the orthonormalized set as $\{ \tilde{e}_\alpha^{pq} \}$. Then the compatible strain tensor closest to $QPP$, and hence which minimizes \eqref{eq:freeenergy}, is
\begin{equation} \label{eq:bestapprox}
	e^\text{best}_\alpha(P) = \sum_{p,q} \langle QPP, \tilde{e}_\alpha^{pq} \rangle_C  \tilde{e}_\alpha^{pq} \, .
\end{equation}
This method only uses simple Fourier analysis and linear algebra, but we have not seen it described elsewhere. The most popular alternative method recasts the elastic contributions as a computationally expensive non-local term \cite{nambu1994domain}.

We can now describe our numerical method. We take some initial data $(P^0, e^0)$ and apply a gradient flow
\begin{align} \label{eq:gradflow}
	\partial_\tau P &= - \frac{\delta F}{\delta P}\bigg\rvert_{(P^0,e^0)} \\
	&= -\left(G_{abcd}\partial_{bc}P_d - \partial_aV + 2q_{\alpha a b}e^0_\alpha P_b\right) \rvert_{P^0}\, .
\end{align}
After each time step, we update the strain tensor using \eqref{eq:bestapprox}. The final solution ($\tau \to \infty$) will be an elastically compatible energy minimizer. Throughout this paper, we use a rectangular grid with approximately $200^2$ grid points and  lattice spacing $0.08$. We stop the gradient flow once the change in energy during the flow has fallen below some critical value.

\begin{figure}[b]
	\begin{center}
    \includegraphics[width=\columnwidth]{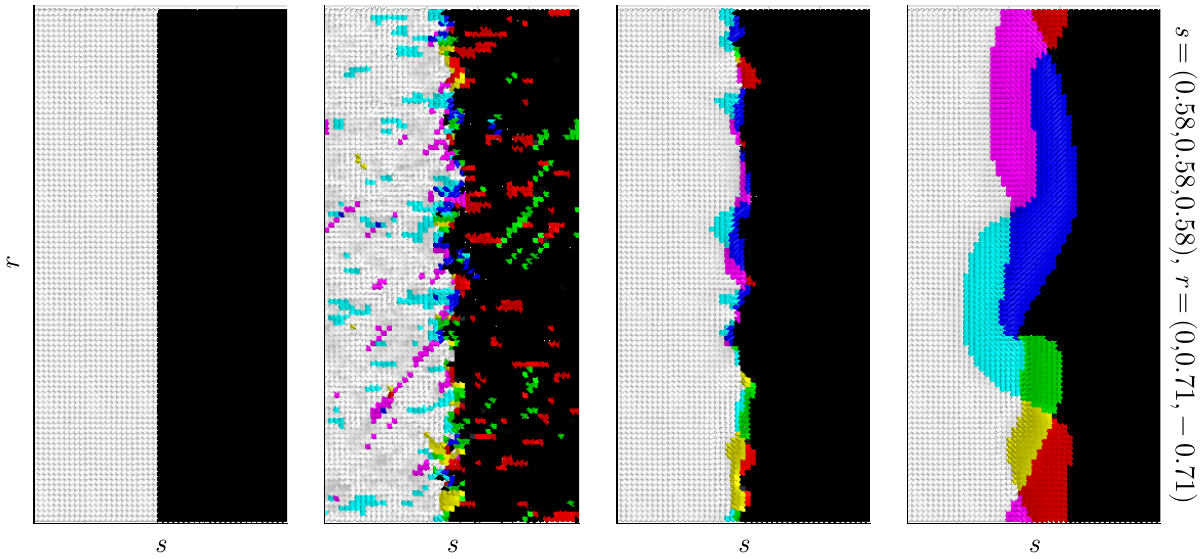}
		\caption{Annealing applied to a system with an Ising domain wall as an initial configuration. We plot the system for times $\tau = 0, 0.08,0.8$ and $80$ from left to right. The final configuration has skyrmion charge $+1$. }
		\label{fig:anneal}
	\end{center}
\end{figure}

To test our claims of a skyrme line appearing in an annealed sample, we have simulated annealing. We take a perioidic sample with $\boldsymbol{s} = (1,1,1)$ and $\boldsymbol{r} = (0,1,-1)$, with a simple Ising domain wall as initial data. We then evolve the system but with random fluctuations. The size of these normally-distributed fluctuations decreases exponentially with time. We plot one simulation in Fig. \ref{fig:anneal} at times $\tau=0,0.08,0.8$ and $80$. We see that the initial Ising Lines develop the structure of a skyrme line. This happened in around $20\%$ of our simulations. This demonstrates that a sample that supports Ising lines is also likely to support skyrme lines after annealing.

\end{document}